\documentclass[aps,physrev,reprint,superscriptaddress,showkeys,amsmath,amssymb]{revtex4-2}

\usepackage{upgreek}
\usepackage{graphicx}

\begin{document}


\title{Integrated optical isolators enabled by 2D semiconductors via spin-momentum locking}


\author{Fernando Martín-Romero}
\email[Contact author: ]{fmarrom@ntc.upv.es}
\affiliation{Nanophotonics Technology Center, Universitat Politècnica de València, Valencia, 46022, Spain}

\author{Víctor J. Gómez}
\affiliation{Nanophotonics Technology Center, Universitat Politècnica de València, Valencia, 46022, Spain}

\author{Alberto G. Curto}
\affiliation{Photonics Research Group, Ghent University-imec, Ghent, 9052, Belgium}
\affiliation{Center for Nano- and Biophotonics, Ghent University, 9052 Ghent, Belgium}


\date{\today}

\begin{abstract}
Integrated optical isolators based on spin-momentum locking hold great potential for protecting photonic integrated circuits from unwanted reflections. Currently, the most common implementations are magneto-optical isolators that require an externally applied magnetic field. Nevertheless, a wide range of nonreciprocal effects remains to be exploited in alternative material platforms. Although the majority of these effects induce changes to the propagation of light as a function of its polarization, they can be directly connected to the propagation direction through spin-momentum locking. We employ a computational approach to investigate various designs of integrated isolators based on nonreciprocal 2D semiconductors, studying the influence of their geometry and optical properties on the device performance. Additionally, we propose a novel concept for an all-optical isolator based on spin–valley polarization and optically tunable magnetism in transition metal dichalcogenide monolayers. Our results establish 2D semiconductors as an advanced platform for integrated nonreciprocal devices, enabling functionalities that are difficult to achieve with conventional magneto-optical materials like all-optical operation.

\end{abstract}


\maketitle

\section{Introduction}

Spin-momentum locking establishes a distinct pathway for the realization of integrated optical isolators by coupling the polarization state of guided light with its propagation direction. Optical isolators are nonreciprocal devices that allow the transmission of light in one direction while blocking it in the opposite. This type of nonreciprocal behavior is highly relevant in photonic integrated circuits (PICs). It protects them from back reflections that can cause interference resulting in  undesired Fabry-Pérot effects and laser instability \cite{jalasWhatWhatNot2013}. Therefore, integrated isolators are a key component of efficient and scalable PICs. In the case of quantum photonic integrated circuits, single-photon nonreciprocity has been demonstrated \cite{renSinglePhotonNonreciprocityIntegrated2022} and further applications have been proposed such as nonreciprocal quantum transfer \cite{mascarenhasNonreciprocalQuantumTransport2019}, or quantum-limited amplification \cite{metelmannNonreciprocalPhotonTransmission2015}.

The objective of any integrated nonreciprocal device is to break the time-reversal symmetry of a photonic circuit, which results in a contrast on the optical response of light propagating in opposite directions. Traditionally, free-space optical isolators have exploited magneto-optic (MO) effects such as Faraday rotation, where the polarization direction of a linearly polarized beam is rotated in the presence of a magnetic field. This rotation originates from a difference in phase velocity between right and left circularly polarized light (RCP and LCP), usually induced in a magneto-optical garnet. The fabrication of integrated isolators based on magnetized garnets is, however, a challenging task due to the lattice mismatch with standard silicon-on-insulator (SOI) substrates \cite{stadlerIntegratedMagnetoOpticalMaterials2014}. Early designs of integrated MO isolators consisted of nonreciprocal TE-TM mode converters based on Faraday rotation followed by polarization filters \cite{warnerFaradayOpticalIsolator1973, andoWaveguideMagnetoopticIsolator1988}. Novel MO isolators generally take advantage of the nonreciprocal phase shift induced by an yttrium iron garnet (YIG) layer \cite{biOnchipOpticalIsolation2011, shojiMagnetoopticalNonreciprocalDevices2014a, yanWaveguideintegratedHighperformanceMagnetooptical2020a}. Alternative designs based on graphene have also been proposed \cite{maHybridGrapheneSilicon2016}, due to its large MO effect \cite{crasseeGiantFaradayRotation2011a}. Nevertheless, approaches based on magneto-optical properties generally present the strong disadvantage of requiring an externally applied magnetic field in order to function. To overcome this drawback, photonic structures based on different effects are being explored. For instance, hybrid graphene-silicon photonic crystal isolators based on nonlinear effects \cite{zareiTunableOpticalIsolator2021} have been proposed, or traveling-wave electro-optic isolators on thin-film lithium niobate \cite{yuIntegratedElectroopticIsolator2023} have been demonstrated.

In this context, spin-momentum locking emerges as a powerful mechanism to bring a wider range of polarization-dependent nonreciprocal effects into integrated photonics. By connecting the transverse photonic spin of a guided mode (and, in turn, its polarization state) to its propagation direction, it provides a direct route to break time-reversal symmetry in the presence of a nonreciprocal material. This strategy allows extending the design of integrated isolators to a broad class of emerging material platforms, such as 2D magnets \cite{zhangAllopticalSwitchingMagnetization2022} or transition metal dichalcogenides (TMDs) \cite{haoOpticallyControllableMagnetism2022}. TMDs are 2D materials consisting of three atomic planes and two atomic species: a transition metal (e.g., W or Mo) and a chalcogen (e.g., S or Se). They present a hexagonal lattice structure, similar to that of graphene. In their Brillouin zone, there exist two different energy band gaps known as K-valley and K'-valley \cite{h.godiksenImpactIndirectTransitions2022, plechingerTrionFineStructure2016a}. If the TMD is optically pumped using circularly polarized light, the pump handedness determines which valley absorbs it, inducing the generation of excitons. In turn, the excitons on each valley exhibit a distinct spin orientation. Therefore, pumping with RCP or LCP light leads to a spin population imbalance, resulting on the magnetization of the TMD.

This optically controllable magnetism enables the emergence of magneto-optic effects without the need of an externally applied magnetic field \cite{haoOpticallyControllableMagnetism2022}. However, to the best of our knowledge, this all-optical nonreciprocity has not yet been exploited in integrated isolators.

Building on this perspective, we design integrated optical isolators based on the spin-momentum locking effect in hybrid TMD-SiN waveguides. We examine how the relevant geometrical and material properties affect the nonreciprocal response. First, by employing a computational approach, we investigate nonreciprocal effects in guided modes. Then, we propose various designs of integrated isolators based on nonreciprocal 2D semiconductors such as absorption-based waveguide isolators and microring resonators, studying the influence of their geometry and optical properties on the device performance. Finally, we discuss the design of Mach-Zehnder integrated isolators based on nonreciprocal phase shift. Together, these configurations provide a general approach to the design of integrated nonreciprocal devices across a range of TMDs and configurations. All in all, these results put forward a novel concept for an all-optical isolator based on spin–valley polarization and optically tunable magnetism in TMD monolayers. Moreover, the isolator design principles we develop are applicable to other emerging materials such as 2D magnets.

\section{Nonreciprocity via spin-momentum locking}

There are two fundamental types of materials that show a differential response to RCP and LCP light: chiral and magneto-optic materials. Chiral materials can be modeled by introducing a magneto-electric coupling through the Pasteur parameter, $\kappa$, such that the constitutive relations can be written as $\bar{D}=\varepsilon\varepsilon_0 \bar{E}-i\kappa\sqrt{\varepsilon_0\mu_0}\bar{H}$ and $\bar{B}=\mu\mu_0\bar{H}+i\kappa\sqrt{\varepsilon_0\mu_0}\bar{E}$ \cite{mohammadi_nanophotonic_2023}. In a typical magneto-optic material with losses, however, the constitutive relations do not incorporate a Pasteur parameter. In turn, the permittivity consists of a complex-valued non-diagonal permittivity tensor, $\bar{\bar{\varepsilon}}$.
Assuming that the applied magnetic field is oriented on the $y$ direction, the diagonal component, $\varepsilon_{xx}$, corresponds to the permittivity of the material in the absence of a magnetic field. The off-diagonal component, $\varepsilon_{xz}$, determines the intrinsic strength of the MO effects in the material. If the MO material is two-dimensional and lies in the $xz$-plane, then $\varepsilon_{yy}$ corresponds to an out-of-plane permittivity component. In the specific case of an optically pumped TMD, the $y$-axis would also correspond to the orientation of the magnetization induced by the pump beam. For the discussed geometry, the relative permittivity tensor is:

\begin{equation}
\bar{\bar{\varepsilon}}=
    \begin{pmatrix}
        \varepsilon_{xx} & 0 & \varepsilon_{xz} \\ 0 & \varepsilon_{yy} & 0 \\ -\varepsilon_{xz} & 0 & \varepsilon_{xx}  \label{eq:tensor}
    \end{pmatrix}
\end{equation}

Both the imaginary parts of $i\kappa$ or $\varepsilon_{xz}$ are responsible for a differential phase velocity between RCP and LCP light, usually defined in free space as optical rotation. Meanwhile, the real parts of $i\kappa$ or $\varepsilon_{xz}$ are responsible for a differential absorption, denoted as circular dichroism (CD). The main difference between chiral and magneto-optic materials arises when the propagation direction of light is reversed. While a chiral material typically has a reciprocal behavior, a magneto-optic material is nonreciprocal. The presence of the external magnetic field (or polarized optical pumping) establishes a preferred direction in space that breaks the time-reversal symmetry, which makes the material suitable for optical isolation. This nonreciprocity is manifested in the opposite signs for both off-diagonal terms of the permittivity tensor, $\varepsilon_{zx}=-\varepsilon_{xz}$.

Since the aim of this work is to study nonreciprocal systems, the focus is on the second family of materials. The eigenstates of the tensor shown in Eq. (\ref{eq:tensor}) correspond to RCP and LCP light, with a permittivity of $\varepsilon_+=\varepsilon_{xx}-i\varepsilon_{xz}$ and $\varepsilon_-=\varepsilon_{xx}+i\varepsilon_{xz}$, respectively. Since $\varepsilon_{xx}$ and $\varepsilon_{xz}$ are complex numbers, this leads to a contrast on the real and imaginary refractive index for circularly polarized light with opposite handedness. The contrast on the real part induces a different phase velocity (Faraday rotation). Meanwhile, the contrast in the imaginary part causes a different absorption coefficient (CD).

This polarization dependence can be conveniently understood in terms of optical spin angular momentum. In free space, a circularly or elliptically polarized plane wave presents a spin angular momentum whose sign is determined by its handedness. This spin is longitudinal, parallel to the propagation direction of the plane wave. In the presence of a MO material, the optical response depends on the relative orientation between the spin angular momentum of light and the magnetization of the material. As a result, opposite spin orientations of light lead to different optical responses, giving rise to Faraday rotation and CD.

In photonic integrated circuits, this spin dependence can be exploited through an effect known as spin-momentum locking \cite{bliokhSpinOrbitInteractions2015, lodahlChiralQuantumOptics2017a}, which establishes a correspondence between the spin angular momentum density of the guided mode and its propagation direction. In the case of a confined transverse electric (TE) mode propagating through a waveguide, the electric field comprises a dominant transverse component and a longitudinal component. These components are phase-shifted by $\pi/2$, leading to an elliptical polarization parallel to the plane of the photonic circuit, and giving rise to a transverse electrical component of the spin angular momentum density. Due to the spatial distribution of the longitudinal and transverse electric field components in a TE$_0$ mode, the orientation of the transverse spin density is opposite on the left and right sides of the waveguide. Upon reversing the propagation direction, the transverse spin distribution changes sign throughout the waveguide, which directly links the propagation direction to the spin orientation \cite{shreinerElectricallyControllableChirality2022b, gongNanoscaleChiralValleyphoton2018, espinosa-soriaTransverseSpinSpinOrbit2016, bliokh_extraordinary_2014a}. A detailed discussion regarding the polarization and spin density distribution of guided modes is provided in the Supplemental Material, Section S1 \cite{supplemental}.

If the guided mode is in the presence of a MO material, the spin-momentum locking enables nonreciprocal behavior due to the preferred spatial direction established by the magnetization direction of the material. In the case of a TE$_0$ mode, the magneto-optic effect must be excited on only one side of the waveguide, so that the transverse optical spin has a well-defined relative orientation with respect to the magnetization. This leads to a contrast in the effective index for light propagating through the waveguide in the forward or backward direction \cite{maHybridGrapheneSilicon2016}. For the real part of the effective index, it induces a nonreciprocal phase shift as a result of the direction-dependent phase velocity. For the imaginary part of the effective index, it induces a nonreciprocal loss due to the direction-dependent absorption coefficient. In the specific case of a TMD, nonreciprocity can be achieved by partially covering one side of the waveguide with a transferred nanoribbon and optically pumping it, either by illuminating the circuit from above with a circularly polarized pump beam, or by waveguide-coupled optical pumping.

\section{Nonreciprocal effects in guided modes}

\begin{figure}[ht!]
\includegraphics{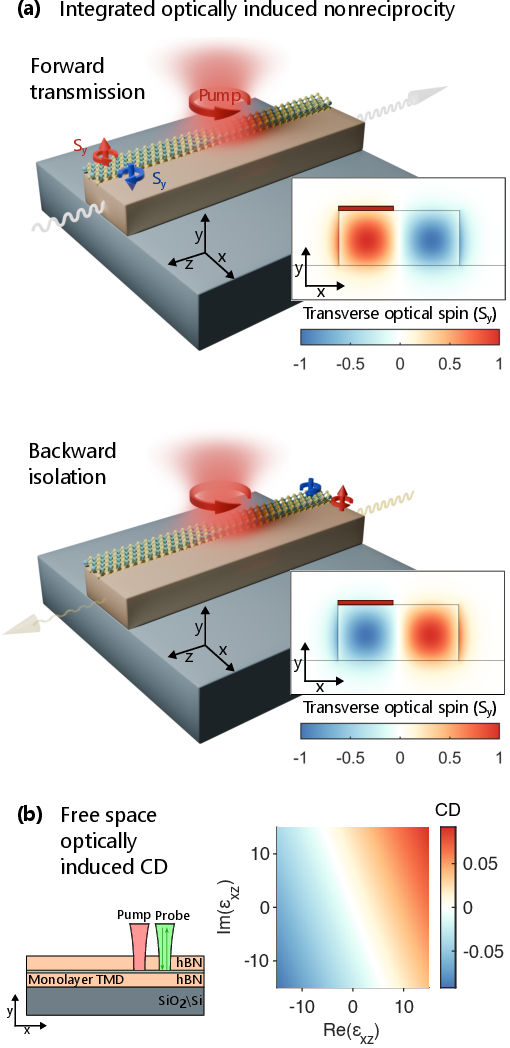}
\caption{\label{fig:1} Nonreciprocal effects in integrated and free-space configurations. (a) Schematic view of a TE single-mode waveguide partially covered by an optically pumped nonreciprocal 2D semiconductor such as a monolayer TMD. In the forward direction, the normalized transverse optical spin at the nonreciprocal 2D semiconductor position is parallel to the pump beam, resulting in transmission. In the backward direction, the spin is opposite and light is absorbed. (b) Free-space circular dichroism (CD) for reflected light in an optically pumped encapsulated TMD monolayer with a diagonal permittivity of $\varepsilon_{xx}=15+15i$.
}
\end{figure}

A TE single-mode waveguide partially covered by an optically pumped TMD monolayer is shown in Fig. \ref{fig:1}(a). Since it covers only one side of the waveguide, the transverse spin density distribution of the TE$_0$ mode has a defined orientation at the position of the TMD . That spin orientation changes sign when the propagation direction is reversed. In the propagation direction in which the orientation of the transverse optical spin density matches the spin angular momentum of the circularly polarized pump, the absorption of the TE$_0$ mode is weaker due to the existing pump-induced exciton population. For clarity, the direction with higher transmission is defined as the forward direction, while the opposite direction is defined as backward.

The nonreciprocity of the described waveguide manifests as a contrast in the effective index for the light propagating through the waveguide in opposite directions. As discussed, this contrast gives rise to direction-dependent effects that can be related to the polarization-dependent effects that would emerge in free-space optics. In this context, the nonreciprocal phase shift (NRPS) is analogous to Faraday rotation, while the nonreciprocal loss (NRL) is analogous to CD. Figure \ref{fig:1}(b) shows the theoretical free-space CD of an optically pumped TMD monolayer encapsulated in hexagonal boron nitride (hBN), as a function of its off-diagonal permittivity. The CD has been computed in reflection using the transfer matrix method at a wavelength of $\lambda=612$ nm, close to the neutral exciton resonance of WS$_2$, using the following definition:

\begin{equation}
    \mathrm{CD}=\frac{R_+-R_-}{R_{0}},
\end{equation}
where $R_\pm$ and $R_0$ correspond to the reflectance of the structure for circularly polarized light under optical pumping or in its absence, respectively.

The chosen range of real and imaginary off-diagonal permittivity values, from 0 to 15, represents an intentionally conservative order of magnitude for a TMD. Within this range, the calculated CD remains under 0.1, well below the values reported experimentally for encapsulated monolayer WSe$_2$, which can reach 0.5 using the same CD definition \cite{haoOpticallyControllableMagnetism2022}. In addition, real and imaginary off-diagonal permittivity values as high as $\approx100$ have been reported for WSe$_2$ and MoSe$_2$ \cite{careyGiantFaradayRotation2024}. Furthermore, an off-diagonal surface conductivity of $1.92\times10^{-4}$ S has been retrieved for optically pumped monolayer WS$_2$ at the exciton resonance \cite{guddala_alloptical_2021}, which corresponds to a real off-diagonal permittivity of $\mathrm{Re}(\varepsilon_{xz})=11.4$. This ensures that the trends observed in our simulations provide a physically reasonable lower-bound estimate for the predicted nonreciprocal effects, and justifies the choice of this order of magnitude for all subsequent simulations. Rather than aiming to predict the performance of a specific design, our study focuses on identifying the key geometrical parameters and material properties that govern the nonreciprocal response and analyzing their influence. The procedure described in this work and the underlying discussed physical principles are fully consistent with different orders of magnitude for the off-diagonal permittivity, which in practice can depend on the specific material and operating conditions. For instance, a quantitative prediction of the performance of an all-optical isolator would require detailed knowledge of parameters dependent on the implementation case, such as the choice of TMD, doping level, pump configuration, and operating wavelength, among others. Nevertheless, the results presented here can be straightforwardly extended to stronger or weaker material responses while preserving the observed trends.

Using COMSOL Multiphysics, we compute the effective index of the TE$_0$ mode in the forward and backward directions via the finite-element method (FEM). The nonreciprocal 2D semiconductor is modeled as a thin layer with thickness $t=0.62$ nm, corresponding to a WS$_2$ monolayer. The nanoribbon width is $w=200$ nm. Its permittivity is described by the tensor in Eq. (\ref{eq:tensor}), where the diagonal element is $\varepsilon_{xx}=15+15i$. This value is well within the order of magnitude of the permittivity near the neutral exciton resonance of WS$_2$ in the absence of pumping, as well as the selected resonant wavelength ($\lambda=612$ nm) \cite{jung_measuring_2019,ermolaev_spectral_2019}. The chosen out-of-plane permittivity, $\varepsilon_{yy}=6.6$, corresponds to the experimentally reported value for WS$_2$ at the same wavelength \cite{munkhbat_optical_2022}. The waveguide consists of a silicon nitride (SiN) core on top of a silicon dioxide (SiO$_2$) bottom cladding. The core dimensions are designed for a single-mode TE waveguide at a wavelength of $\lambda=612$ nm, with a width of $W=450$ nm and a thickness of $H=200$ nm.

\begin{figure*}[ht!]
\includegraphics{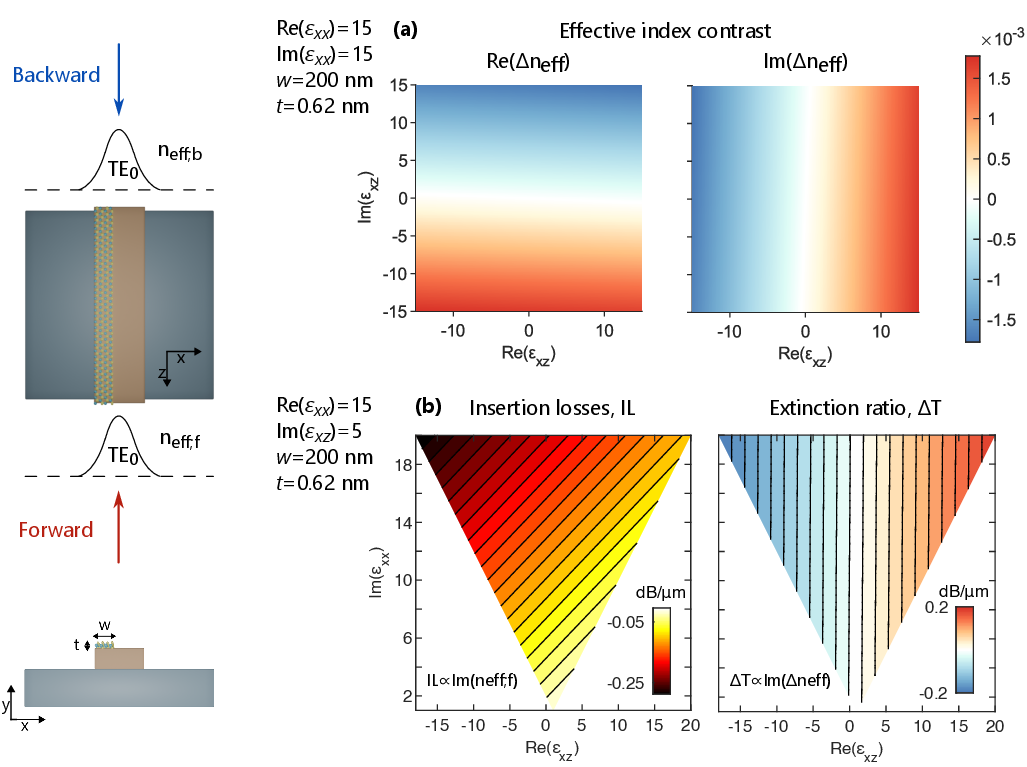}
\caption{\label{fig:2} Integrated nonreciprocal index and attenuation contrast. (a) Real and imaginary effective index contrast between forward and backward propagation, induced for the TE$_0$ mode in a SiN waveguide partially covered by a nonreciprocal monolayer semiconductor. The contrast on the real part depends predominantly on the imaginary part of the off-diagonal permittivity, and vice versa. (b) Insertion losses and extinction ratio derived from the imaginary part of the effective index.
}
\end{figure*}

We performed a parametric study on the real and imaginary part of the off-diagonal permittivity, in order to analyze its contribution on the effective index contrast between both propagation directions [Fig. \ref{fig:2}(a)]. As can be observed, the real effective index contrast, $\mathrm{Re}(\Delta n_{\mathrm{eff}})$, mainly depends on the imaginary part of the off-diagonal permittivity, $\mathrm{Im}(\varepsilon_{xz})$. Meanwhile, the imaginary effective index contrast, $\mathrm{Im}(\Delta n_{\mathrm{eff}})$, mainly depends on the real part of the off-diagonal permittivity, $\mathrm{Re}(\varepsilon_{xz})$. This behavior derives from the definition of the permittivity tensor, for which the real part of the eigenvalues depends on $\mathrm{Im}(\varepsilon_{xz})$, and vice versa:

\begin{subequations}
\begin{align}
    \mathrm{Re}(\varepsilon_{\pm})&=\mathrm{Re}(\varepsilon_{xx})\mp \mathrm{Im}(\varepsilon_{xz}), \label{eq:re_eps_cp}\\
    \mathrm{Im}(\varepsilon_{\pm})&=\mathrm{Im}(\varepsilon_{xx})\mp \mathrm{Re}(\varepsilon_{xz}). \label{eq:im_eps_cp}
\end{align}
\end{subequations}

A variety of designs can be proposed for integrated isolators based on nonreciprocal waveguides. The most straightforward approach consists in an absorption-based isolator, which takes advantage of the contrast in the imaginary effective index (i.e., the nonreciprocal loss, NRL). The absorption coefficient of the TE$_0$ mode in dB per unit length, $\alpha$, is related to its imaginary effective index as

\begin{equation}
   \alpha=-40\pi\cdot\log_{10}\left(\mathrm{e}\right)\left(\frac{\mathrm{Im}(n_{\mathrm{eff}})}{\lambda}\right). \label{eq:att}
\end{equation}
The extinction ratio of the absorption-based isolator has been computed as the transmission contrast between forward and backward propagated light, in units of dB/cm. The insertion losses of the device are defined as the light absorption in the forward direction of the waveguide, which corresponds to the direction of transmission, in units of negative dB/cm. The attenuation of light depends on two contributions. The first is reciprocal absorption, which is dominated by the imaginary permittivity in the absence of optical pumping, $\mathrm{Im}(\varepsilon_{xx})$. The second is nonreciprocal absorption, which is dominated by $\mathrm{Re}(\varepsilon_{xz})$, as extracted from Fig. \ref{fig:2}(a) and Eq. (\ref{eq:att}). The influence of these two quantities was therefore studied, as illustrated in Fig. \ref{fig:2}(b), since they are the key parameters controlling the performance of the absorption-based isolator. In contrast, the real part of the diagonal permittivity was fixed as $\mathrm{Re}(\varepsilon_{xx})=15$ and the imaginary part of the off-diagonal permittivity was fixed as $\mathrm{Im}(\varepsilon_{xz})=5$, taken as a representative intermediate value within the explored range. Both mainly affect the real part of the permittivity eigenvalues [Eq. (\ref{eq:re_eps_cp})], and have negligible impact on light attenuation. Because the extinction ratio arises only from nonreciprocal absorption, it can be observed that it primarily depends on $\mathrm{Re}(\varepsilon_{xz})$. On the other hand, insertion losses in the forward direction are a consequence of the total absorption due to the reciprocal and nonreciprocal contributions, and therefore depend on both $\mathrm{Re}(\varepsilon_{xz})$ and $\mathrm{Im}(\varepsilon_{xx})$. Overall, the results indicate that a high real off-diagonal permittivity and a low diagonal permittivity contribute to a more compact device with lower insertion losses.

The diagonal and off-diagonal permittivities of a nonreciprocal 2D semiconductor may be correlated. Although we vary these permittivity components independently to isolate their individual contributions, this parametric analysis is intended to identify the role of each tensor component.
In particular, both the reciprocal and nonreciprocal absorption are expected to be stronger for wavelengths close to the excitonic resonance of a TMD \cite{jung_measuring_2019,ermolaev_spectral_2019, guddala_alloptical_2021}. A specific application case for the framework developed in this work based on experimentally measured off-diagonal surface conductivity in monolayer TMDs is presented as an example in the Supplemental Material, Section S2 \cite{supplemental}.

\section{Absorption-based integrated isolators}

\begin{figure*}[t!]
\includegraphics{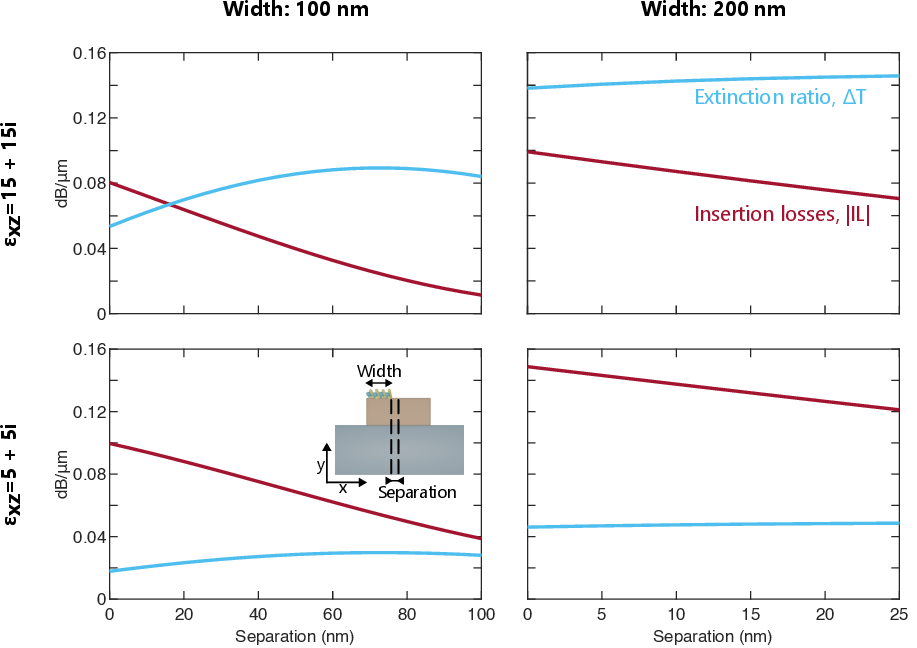}
\caption{\label{fig:3}Extinction ratio and insertion losses for the direct absorption isolator for varying monolayer width and separation with respect to the center of the TE single-mode SiN waveguide, at the wavelength of $612$ nm. It can be observed that the extinction ratio and losses tend to improve with a larger separation. Increasing the width improves the extinction ratio at the expense of higher losses. For materials with a high off-diagonal permittivity (strong nonreciprocal effect, $\varepsilon_{xz}=15+15i$ for the assigned representative TMD), there is a crossing point where the extinction ratio is higher than the losses. But for weaker nonreciprocal materials ($\varepsilon_{xz}=5+5i$ for the representative TMD), such a crossing point can not be reached.
}
\end{figure*}

In the previous section, we established that a silicon nitride waveguide partially covered by a nonreciprocal 2D semiconductor nanoribbon can act as an integrated isolator based on nonreciprocal loss. The influence of the off-diagonal permittivity on the performance of the device has already been discussed. However, the efficiency also depends on the geometry of the nanoribbon. Figure \ref{fig:3} presents the variation of the extinction ratio and insertion losses of the absorption-based isolator with respect to two parameters: the width of the nanoribbon and its horizontal separation with respect to the symmetry plane of the waveguide. Increasing the width leads to a better contrast at the expense of higher losses. Separating the nanoribbon from the center of the waveguide reduces the insertion losses while increasing the extinction ratio. This trend is a consequence of the elliptical polarization state distribution for a TE$_0$ mode, which progresses from a pure in-plane circular polarization near the edges of the SiN core, to a linear polarization at the symmetry plane of the waveguide (see Fig. S2 within the Supplemental Material \cite{supplemental}). For materials with stronger nonreciprocal effects (e.g. $\varepsilon_{xz}=15+15i$), a separation threshold value can exist above which the extinction ratio exceeds the losses. If the effect is weaker (e.g. $\varepsilon_{xz}=5+5i$), this threshold may not be observed.

\begin{figure*}[ht!]
\includegraphics{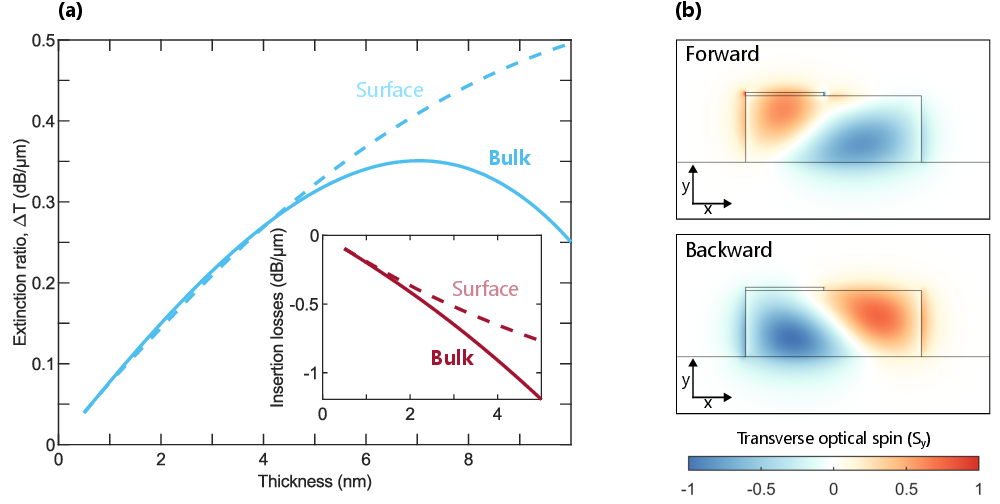}
\caption{\label{fig:4}Effect of the nanoribbon thickness on a nonreciprocal waveguide. (a) Extinction ratio and insertion losses for a direct absorption isolator based on the TE single-mode SiN waveguide at $\lambda=612$ nm partially covered by a 200 nm wide nonreciprocal semiconductor of varying thickness, with $\varepsilon_{xz}=5+5i$. Two models have been used: a 2D material with a surface current or a thin layer with a bulk permittivity. It can be observed that the bulk approach predicts a decrease in the extinction ratio for thicker layers, and that in both cases the losses rapidly increase with thickness. (b) Deformation of the normalized transverse spin profile in the waveguide that causes the decrease in extinction ratio, for a material with $t=10$ nm and $\varepsilon_{xz}=15+15i$.
}
\end{figure*}

To study the influence of the two-dimensional character of the nonreciprocal material on the performance of the device, we calculate the contrast and losses as a function of nanoribbon thickness [Fig. \ref{fig:4}(a)]. For this purpose, two different approaches were followed. On the one hand, the nanoribbon was modeled as a thin layer with the permittivity tensor of Eq. (\ref{eq:tensor}), following the same procedure as in the rest of the simulations presented in this work. On the other hand, it was modeled as a two-dimensional layer with a surface conductivity, $\bar{\bar{\sigma}}$, related to the permittivity through the following equation:

\begin{equation}
    \bar{\bar{\sigma}}=\frac{\omega t\varepsilon_0}{i}(\bar{\bar{\varepsilon}}-\bar{\bar{1}}),
\end{equation}
where $\omega$ is the angular frequency of light and $t$ is the thickness of the nanoribbon. It should be noted that the performance predicted by both models is equal in the case of a monolayer ($t<1$ nm). However, they start to diverge for multilayered nanoribbons. The reason is that the surface current approach starts to lose validity, since it does not take into account the deformation of the transverse spin profile due to the presence of the nanoribbon, which can be clearly visualized in Fig. \ref{fig:4}(b) for a high off-diagonal permittivity. In fact, for a thickness over $7$ nm, the extinction ratio of the isolator starts decreasing because the spin density distribution is no longer antisymmetric. The losses rapidly increase with thickness. These results support the conclusion that a monolayer is the most suitable option for this type of device.

\section{Microring resonators for optical isolation}

The absorption-based isolator concept already demonstrated with simple straight waveguides can be naturally extended to more complex devices. For instance, a resonant cavity can be used to enhance the nonreciprocal effect \cite{maHybridGrapheneSilicon2016}. In the case of a bus waveguide coupled to a microring resonator, the transmittance through the waveguide can be expressed as

\begin{equation}
    T=\frac{\gamma^2-2r\gamma \cos(\phi)+r^2}{1-2r\gamma \cos(\phi)+r^2\gamma^2},
\end{equation}

where $r$ is the amplitude self-coupling coefficient of the bus waveguide, $\phi=2\pi Rk_0 \mathrm{Re}(n_{\mathrm{eff}})$ is the single-pass phase shift, $\gamma=\exp[-2\pi R k_0 \mathrm{Im}(n_{\mathrm{eff}})]$ is the single-pass amplitude transmission through the ring, $R$ is the ring radius and $k_0$ is the wavenumber of light in free space \cite{yarivCriticalCouplingIts2002a}. At resonance, the condition $\phi=2\pi N$ is fulfilled, with $N$ an integer. Thus, the resonant transmittance through the bus waveguide becomes $T=(\gamma-r)^2/(1-\gamma r)^2$, which vanishes under critical coupling conditions, $r=\gamma$.

\begin{figure*}[ht!]
\includegraphics{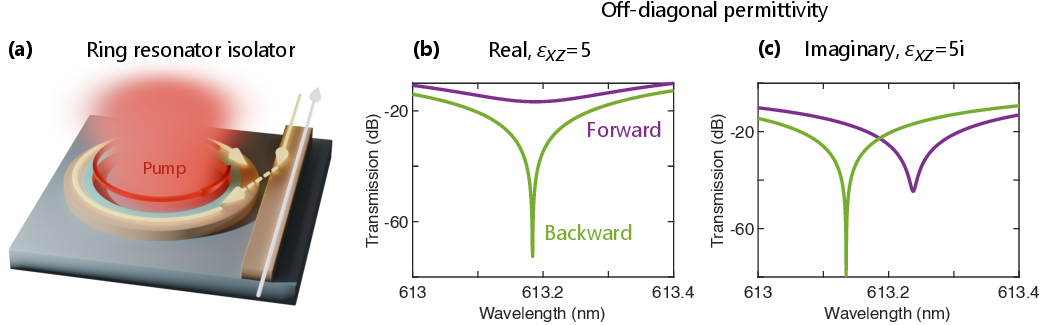}
\caption{\label{fig:5} Ring-resonator-induced nonreciprocity and optical isolation. (a) Schematic view of an isolator based on a waveguide coupled to a ring resonator with radius $R=10$ $\mathrm{\upmu}$m. The ring is partially covered with a $200$ nm wide nonreciprocal 2D semiconductor. (b) Transmission through the waveguide for a real off-diagonal permittivity $\varepsilon_{xz}=5$, assuming the backward-propagating mode fulfills the critical coupling condition and the laser has a narrow bandwidth ($\sim$1 pm). (c) Analogous study for an imaginary off-diagonal permittivity $\varepsilon_{xz}=5i$. It can be observed that the real $\varepsilon_{xz}$ only induces a contrast in the resonance strength, while the imaginary $\varepsilon_{xz}$ also induces a wavelength shift. The structure of the bus waveguide and ring corresponds to the TE single-mode SiN waveguide at $\lambda=612$ nm.}
\end{figure*}

If a nonreciprocal 2D nanoribbon is placed on top of the inner radius of the ring resonator, the resonant and critical coupling conditions become different for clockwise and counter-clockwise propagation [Fig. \ref{fig:5}(a)]. As a result, the transmission through the bus waveguide becomes nonreciprocal, which can be exploited to build an optical isolator. Assume that the gap between the waveguide and the ring resonator is chosen such that the backward-propagating light operates under the critical coupling condition, $r=\gamma_b$. If the nonreciprocal nanoribbon exhibits a real off-diagonal permittivity, the imaginary effective index differs for clockwise and counter-clockwise modes. Consequently, forward-propagating light is mostly transmitted through the bus waveguide, while backward-propagating light is critically coupled into the ring [Fig. \ref{fig:5}(b)]. If, instead, the off-diagonal permittivity is imaginary, a contrast on the real effective index appears. This induces a shift in the resonant wavelength between clockwise and counter-clockwise propagation. If the backward-propagating light is on resonance with the ring, the forward-propagating light is detuned from resonance, leading to waveguide transmission mostly in the forward direction \ref{fig:5}(c). In this second case, the minor contribution of the imaginary off-diagonal permittivity to the change on the imaginary effective index also introduces a contrast in critical coupling, since it is susceptible to slight variations. The purely real and purely imaginary values of $\varepsilon_{xz}$ are considered separately to isolate the underlying physical mechanisms; for a complex $\varepsilon_{xz}$, both effects appear simultaneously as well. 

\section{Nonreciprocal phase shift in Mach-Zehnder isolators}

Nonreciprocal phase shift (NRPS) can also be exploited to design an integrated isolator \cite{shojiMagnetoopticalNonreciprocalDevices2014a}. It can be expressed as a contrast in the phase constant of light propagating in opposite directions, and it depends on the real part of the effective index as

\begin{equation}
    \Delta\beta=k_0\mathrm{Re}(\Delta n_{\mathrm{eff}}).
\end{equation}

In order to build the isolator, the first step is to design a Mach-Zehnder interferometer (MZI) with a reciprocal phase difference of $\pi/2$ between both arms. Then, a nonreciprocal 2D material can be integrated into the device to generate an additional phase difference of $-\pi/2$ in one direction and $\pi/2$ in the opposite. By adding the reciprocal and nonreciprocal contributions, the total phase difference is zero in one direction, leading to constructive interference between both arms. In the opposite direction, this phase difference is $\pi$, resulting in destructive interference.

\begin{figure}[ht!]
\includegraphics{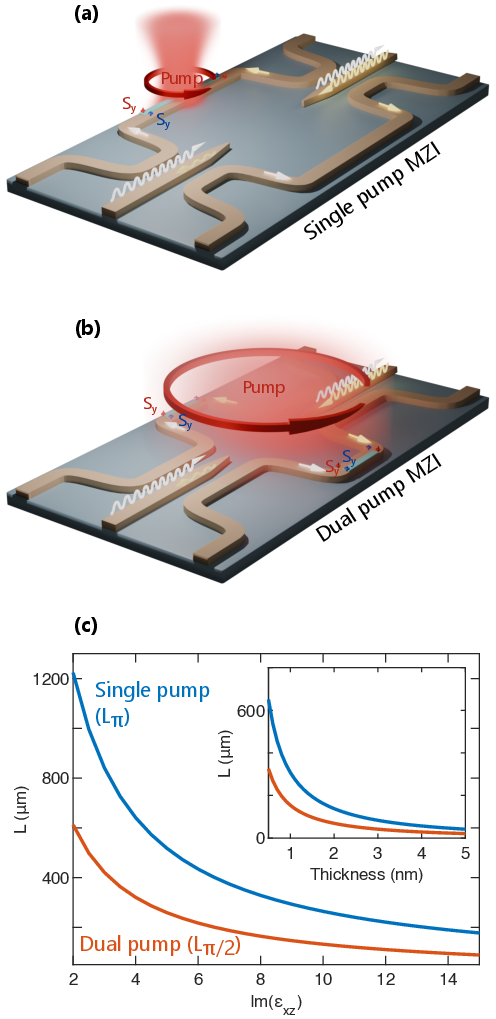}
\caption{\label{fig:6}Nonreciprocal phase shift and optical isolation in Mach-Zehnder interferometers. (a-b) Schematic view of the MZI isolator with single or dual pump, respectively. In the dual-pump configuration, the nonreciprocal monolayer is located on opposite sides for each arm. (c) Propagation length required to build a MZI isolator based on single-arm and dual-arm pumping configurations, as a function of $\mathrm{Im}(\varepsilon_{xz})$ for a nonreciprocal monolayer. Inset: propagation length as a function of the nonreciprocal layer thickness, assuming $\varepsilon_{xz}=5+5i$. The base structure considered for both arms of the MZI corresponds to the TE single-mode SiN waveguide at $\lambda=612$ nm. The nanoribbons are $200$ nm wide and positioned next to the core edge.
}
\end{figure}

To achieve the additional nonreciprocal phase difference of $\pm\pi/2$ between both arms, there are two possible configurations. The first consists of a single pump configuration where one arm of the MZI is partially covered with a nonreciprocal nanoribbon, while the other arm remains a reciprocal waveguide [Fig. \ref{fig:6}(a)]. Then, the additional phase difference between the two arms in the forward and backward directions is, respectively,

\begin{subequations}
\begin{align}
    \mathrm{Re}(n_{\mathrm{F}})k_0 L_1-\mathrm{Re}(n_{\mathrm{eff}})k_0 L_2&=\pi/2, \\
    \mathrm{Re}(n_\mathrm{{B}})k_0 L_1-\mathrm{Re}(n_{\mathrm{eff}})k_0 L_2&=-\pi/2,
\end{align}
\end{subequations}
where $n_{\mathrm{eff}}$, $n_\mathrm{F}$ and $n_\mathrm{B}$ are the effective indices of the TE$_0$ mode in the reciprocal waveguide, and in the forward and backward directions of the nonreciprocal waveguide. The lengths of the nonreciprocal and reciprocal arms are $L_1$ and $L_2$, respectively. Solving this system of equations leads to the following relations:

\begin{subequations}
\begin{align}
    L_1&=\frac{\pi}{\Delta\beta}, \\
    L_2&=\frac{\mathrm{Re}(n_\mathrm{F})+\mathrm{Re}(n_\mathrm{B})}{2\mathrm{Re}(n_{\mathrm{eff}})}L_1.
\end{align}
\end{subequations}

The first equation indicates that the length of the nonreciprocal arm must yield an NRPS of $\pi$. If the other arm was covered by a reciprocal nanoribbon (e.g., an unpumped TMD monolayer) of the same dimensions, its length should be equal to that of the first arm ($L_2=L_1$). However, assuming it is a bare SiN waveguide not covered by a TMD, its length requires a small correction due to the mismatch in the effective index.

The second design option relies on a push-pull dual pump MZI configuration, in which both arms are covered with nonreciprocal nanoribbons but placed on opposite sides [Fig. \ref{fig:6}(b)]. Since the transverse spin density of the TE$_0$ mode has opposite orientations on each side of the waveguides, the resulting nonreciprocal effect in the two arms is opposite. Therefore, the additional phase difference in the forward and backward propagation directions is, respectively,

\begin{subequations}
\begin{align}
    \mathrm{Re}(n_{F})k_0 L_1-\mathrm{Re}(n_{B})k_0 L_2&=\pi/2, \\
    \mathrm{Re}(n_{B})k_0 L_1-\mathrm{Re}(n_{F})k_0 L_2&=-\pi/2.
\end{align}
\end{subequations}

This system leads to the following result:

\begin{equation}
    L_1=L_2=\frac{\pi/2}{\Delta\beta},
\end{equation}
which indicates that both nonreciprocal arms must have a length that yields an NRPS of $\pi/2$. Consequently, the push-pull configuration results in a device with half the length of the single-pump configuration. Nevertheless, the operation of the isolator becomes more complex. For instance, if the nanoribbons are composed of a TMD, optical pumping would be required in both arms. Figure \ref{fig:6}(c) shows the computed operation length of both MZI isolators as a function of the imaginary off-diagonal permittivity, as well as the nanoribbon thickness using the thin-layer permittivity method. The device footprint becomes more compact for larger imaginary off-diagonal permittivity, since it is the material parameter that governs the NRPS effect. Although the MZI isolator presents a larger footprint, it can theoretically achieve a perfect extinction ratio due to destructive interference.

\section{Conclusions}

We have provided a detailed framework for the design and simulation of nonreciprocal waveguides and integrated isolators based on 2D semiconductor nanoribbons and spin-momentum locking, applicable to a broad range of materials including 2D magnets and TMDs. We considered two different models for the nanoribbons: a thin layer with an off-diagonal permittivity and a surface off-diagonal conductivity. While both methods are equally valid in the case of a monolayer, only the permittivity model produces realistic results for multilayered nanoribbons. We discussed the geometry and optical properties that the nonreciprocal 2D semiconductor nanoribbons must fulfill. A low imaginary diagonal permittivity is desirable in any case to avoid insertion losses in the waveguide. In addition, a large imaginary off-diagonal permittivity favors NRPS, whereas a high real off-diagonal permittivity results in greater NRL with lower absolute insertion losses.

If the 2D semiconductor exhibits a strong NRL, it is suitable for the design of absorption-based isolators consisting of a simple waveguide partially covered by a nanoribbon of the material. If the NRPS is more pronounced, then it can be effectively applied to design nonreciprocal Mach-Zehnder interferometers. Any combination of these two conditions can be useful in a microring isolator, where the coupling conditions between the bus waveguide and the ring are influenced by both effects. Although the ring resonator presents a higher extinction ratio and a more compact design than the waveguide, it also leads to larger insertion losses and a narrower bandwidth. The MZI has the largest footprint but achieves a perfect isolation ratio due to destructive interference.

Beyond these design considerations, we have proposed optically pumped isolators implemented with transition metal dichalcogenides. This novel approach, based on spin-valley polarization in this family of materials, enables all-optical nonreciprocity and paves the way for the realization of photonic integrated circuits incorporating optical isolators.


\acknowledgments
We thank Ankit Singh and Tom Sistermans for discussions about the simulation methods. We acknowledge the financial support by the Generalitat Valenciana (CIACIF/2022/188, CIBEFP/2023/108 and ESGENT/2024/17), the AGENCIA ESTATAL DE INVESTIGACIÓN of Ministerio de Ciencia, Innovacion y Universidades (PID2020-118855RB-I00, PID2024-162261NB-I00 and Grant CNS2023-145093 funded by MICIU/AEI/10.13039/501100011033 and by “European Union NextGenerationEU/PRTR”), and partly by the Flemish Research Foundation (FWO) under grant G066924N and the European Research Council (ERC) under the European Union's Horizon 2020 Research and Innovation Program (Grant Agreement 948804, CHANSON).

\bibliography{bibliography}

\end{document}